\documentclass{elsart}

\usepackage{graphicx}

\usepackage{amssymb}

\begin{document}

\begin{frontmatter}


\title{Stable regions and singular trajectories in chaotic soft wall billiards}
\author{Ariel Kaplan, Nir Friedman, Mikkel Andersen, and Nir Davidson}

\address{Department of Physics of Complex Systems, Weizmann Institute of Science, Rehovot 76100, Israel }

\begin{abstract}
We present numerical and experimental results for the development
of islands of stability in atom-optics billiards with soft walls.
As the walls are soften, stable regions appear near singular
periodic trajectories in converging (focusing) and dispersing
billiards, and are surrounded by areas of "stickiness" in
phase-space. The size of these islands depends on the softness of
the potential in a very sensitive way.
\end{abstract}

\end{frontmatter}

\section{Introduction}
\label{} Kolmogorov-Arnold-Moser (KAM) theory provides a framework
for the understanding of an integrable Hamiltonian system which is
subject to a perturbation \cite{Gutzwiller90}. As the strength of
the perturbation is increased areas of phase-space became chaotic
and eventually only islands of stability, surrounded by a chaotic
"sea", survive. The opposite question is of interest as well: What
will happen to a completely chaotic system when a perturbation is
applied? will it remain chaotic or will islands of stability
appear? The appearance of KAM islands is of great importance:
First, for initial conditions falling inside the island, there
will be finite range oscillations in the the tail of the
correlation function. In addition, these islands are usually
accompanied by areas of "stickiness" where particles belonging to
the chaotic region of phase space spend an anomalous amount of
time, and hence they affect the temporal correlation function even
for initial conditions in the chaotic "sea". Indeed, it was argued
that this stickiness can cause a power-law decay of correlations,
instead of an exponential one, and thus affect the transport
properties of the whole system even for a relatively small island
size\cite{Zaslavsky99}.

For the billiard system (i.e. a particle moving freely in a
bounded region and reflecting elastically from the boundary), one
such perturbation is making the walls "soft", as opposed to the
infinitely steep potential of the ideal case. Since the billiard
is a widely used paradigm for understanding the microscopic
foundations of statistical mechanics, and noting that real
classical molecules move in smooth potentials, we conclude that
the study of soft-wall billiards is an important step towards
sustaining these foundations and, in particular, Boltzmann's
ergodic hypothesis.

For a certain kind of dispersing billiards it was theoretically
proven\cite{Turaev98,Romkedar99} that when the wall becomes soft,
an island appears near a singular, tangent trajectory (see details
in section 3). More recently, and partly inspired by the numerical
results presented herein, the appearance of an additional island,
this time near a corner, was also shown\cite{Turaev02}.
Experimentally, some conjectures were made about the role played
by the softness of walls in semi-conductor quantum dots. It was
considered to be responsible for the algebraic tail in the
distribution of the length of trajectories, which gives rise to
the fractal nature of conductance
fluctuations\cite{Ketzmerick96,Sachrajda98,Ouchterlony99,Sachrajda99}.
Observations of the formation of stable islands in a stadium
shaped soft atom-optics billiard were reported in Ref.
\cite{Kaplan02}.

In this article, we present additional results of numerical and
experimental studies of soft wall atom-optics billiards. Three
unifying concepts are found in our results: First, islands of
stability arise around singular trajectories (tangencies, corners)
as predicted by Refs. \cite{Turaev98,Romkedar99,Turaev02}. Next,
the size of these islands depend on the wall's softness in a very
sensitive way. Finally, "stickiness" regions surround the stable
islands and affect the decay properties of the billiards. In
section 2, we present our numerical and experimental studies of
the stability induced in an atom-optics billiard of the Bunimovich
type\cite{Bunimovich79} by making the walls soft. In section 3, we
present results of numerical simulations for the dynamics of atoms
inside totally dispersing billiards, which exhibit softness
induced stability around singular trajectories.

\section{Islands of stability in converging billiards}

\begin{figure}[tbp]
\begin{center}
\includegraphics[width=5in]{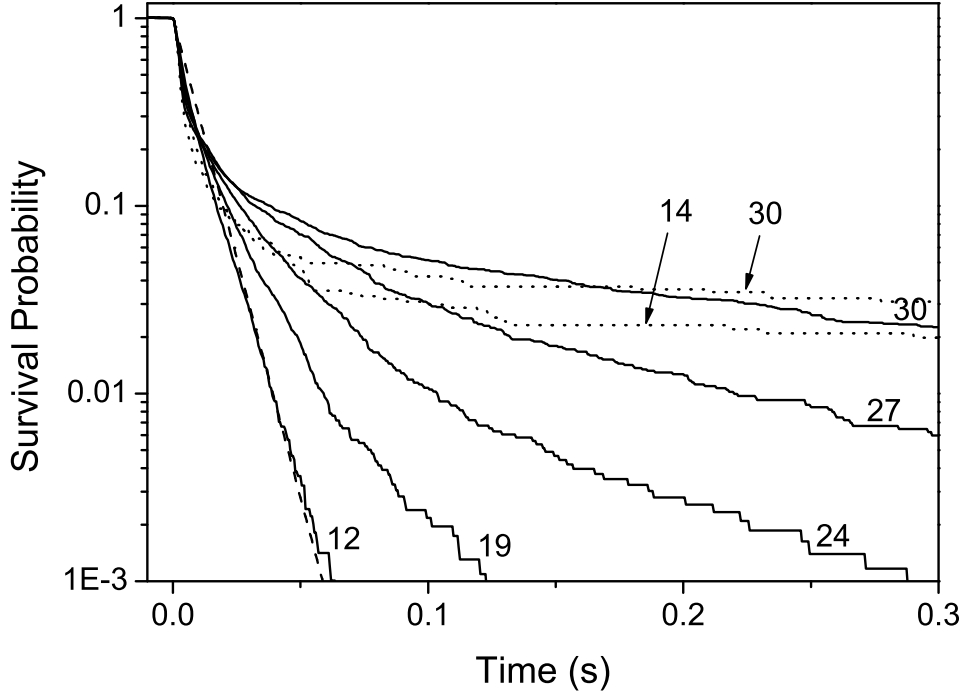}
\end{center}
\caption{Numerical simulations for the decay, through a small
hole, from a tilted Bunimovich stadium (solid lines) and a
circular billiard (dotted lines) with the parameters specified in
the text, for several values of the softness parameter. Both
billiards have the same area and hole size. The numbers adjacent
to the lines indicate the value (in microns) of the softness
parameter $w_{0}$. The simulations for the stadium show a
monotonic slowing down in the decay and increase in stability as
the walls are made softer. For the circular billiard, there is
almost no change in the decay curve as a consequence of the higher
softness. The dashed line shows exp(-t/$\protect\tau _{c} $),
where $\protect\tau _{c}$ is the escape time calculated for the
experimental parameters. } \label{simstad}
\end{figure}

The atom-optics billiard is a recently developed experimental
system in which billiard dynamics can be
studied\cite{Milner01,Friedman01}. As described in detail in Ref.
\cite{Friedman01}, the billiard is realized by the use of a laser
beam which is rapidly ($100$ kHz) scanned using two perpendicular
acousto-optic scanners (AOSs). The laser frequency is tuned above
the atomic resonance (the $D_{2}$ line of $^{85}$Rb in our
experiment), hence it applies a repulsive force on the atoms. By
controlling the deflection angle of both AOSs, we create the
required billiard shapes which confine the atoms in the transverse
direction. The instantaneous potential is given by the dipole
potential of the laser gaussian beam:

\begin{equation}
U\left( x,y,t\right) =U_{0}\exp \left[ -2\left( \left(
x-x_{0}\left( t\right) \right) ^{2}+\left( y-y_{0}\left( t\right)
\right) ^{2}\right) /w_{0}^{2}\right],
\end{equation}
where the curve $(x_{0}\left( t\right) ,y_{0}\left( t\right) )$ is
the shape of the ideal billiard, along which the center of the
Gaussian laser beam scans. $w_{0}$ is the laser beam waist and
$U_{0}$ is the instantaneous potential height. The time averaged
potential is kept constant for different values of $w_{0}$ by
changing the laser power. In such way, $w_{0}$ serves as the
softness control parameter, and is experimentally controlled by
the use of a a telescope with a variable magnification, located
prior to the AOS's such that $w_{0}$ can be changed without
affecting the billiard's size and shape. Fast enough scanning of
the beam results in an effective time-averaged potential wall
\cite{friedman00}. Photon scattering from the laser beam is
greatly suppressed due to the large detuning, and collisions
between the atoms are very rare for the range of densities used,
hence motion of the atoms between reflections from the light walls
can be regarded as strictly ballistic\cite{Friedman01}.

\begin{figure}[tbp]
\begin{center}
\includegraphics[width=5in]{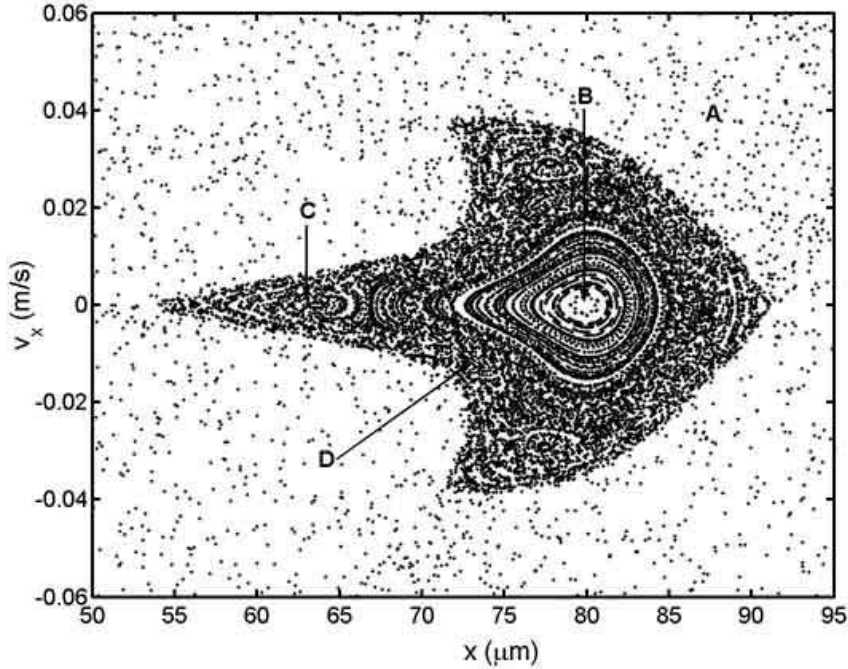}
\end{center}
\caption{Poincar\'{e} surface of section for monoenergetic atoms
confined in a soft ($w_{0}=24\protect\mu $m) tilted-stadium shaped
atom-optics billiard with parameters as in Fig. 1. The Figure
shows $\sim 9\%$ of the total phase space area. A phase space area
corresponding to chaotic trajectories is marked "A", one of period
2 trajectories inside an island "B", an area of higher periodicity
trajectories "C" and one of "sticky" trajectories "D". }
\label{psstad}
\end{figure}

The loading scheme of cold atoms into the billiard was described
in \cite{Kaplan02}. Briefly, laser cooled $^{85}$Rb atoms are
loaded from a magneto optical trap into a red-detuned
one-dimensional optical lattice which is laterally shifted (250
$\mu $m) from the billiard's location. The atoms are then
transferred from the lattice into the billiard by pushing them
with a pulse of a strong on-resonance beam which is perpendicular
to both the lattice and the billiard beams. Simultaneously with
the pushing, the lattice beams are adiabatically switched-off,
resulting in additional cooling\cite {Winoto99}. Further reduction
of the radial velocity spread and especially a decrease in the
number of very slow atoms is achieved by capturing only a central
velocity group of atoms into the billiard, which is turned on at
the proper time after the push. Typically, $3\cdot 10^{5}$ cold
$^{85}$Rb atoms are loaded into the billiard, with a typical
velocity of $\sim 100$ mm/s and a velocity spread of $\sim 20$
mm/s. The cooling in the longitudinal direction ensures that the
system can be approximated as a two dimensional one.

\begin{figure}[tbp]
\begin{center}
\includegraphics[width=5in]{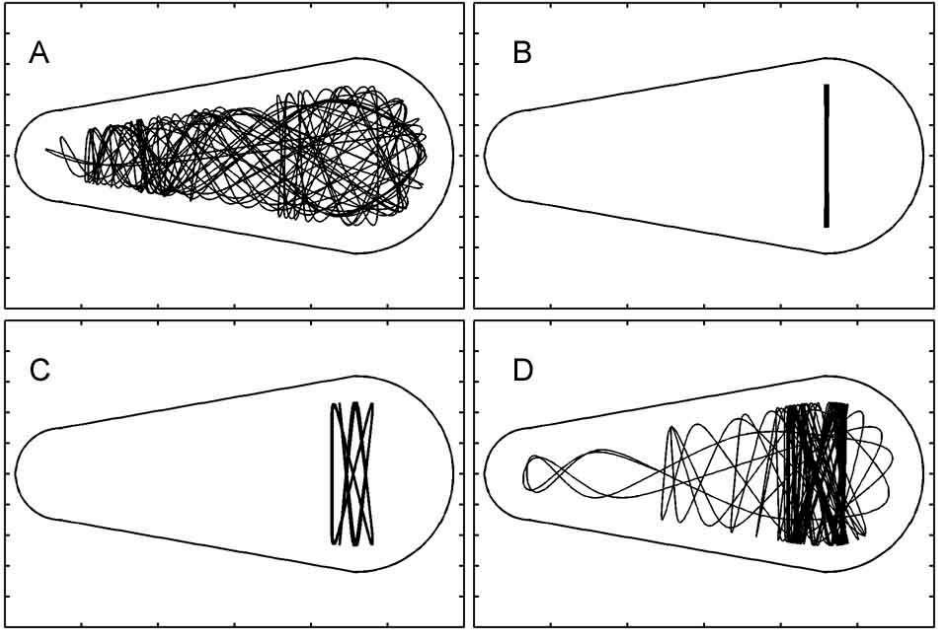}
\end{center}
\caption{Typical trajectories for atoms confined in a soft
($w_{0}=24\protect\mu $m) tilted-stadium shaped atom-optics
billiard with parameters as in Fig. 1. (A) A chaotic trajectory,
corresponding to the area marked "A" in Fig. \ref{psstad}. (B) A
periodic trajectory of period 2 and belonging to the central
island, marked "B" in Fig. \ref{psstad}. (C) A high periodicity
trajectory (Smaller islands surrounding the central one and
denoted "C" in Fig. \ref{psstad}). (D) A "Sticky" trajectory,
spending in the island's vicinity a relative long time ("D" in
Fig. \ref{psstad}).} \label{trajstad}
\end{figure}

\begin{figure}[tbp]
\begin{center}
\includegraphics[width=5in]{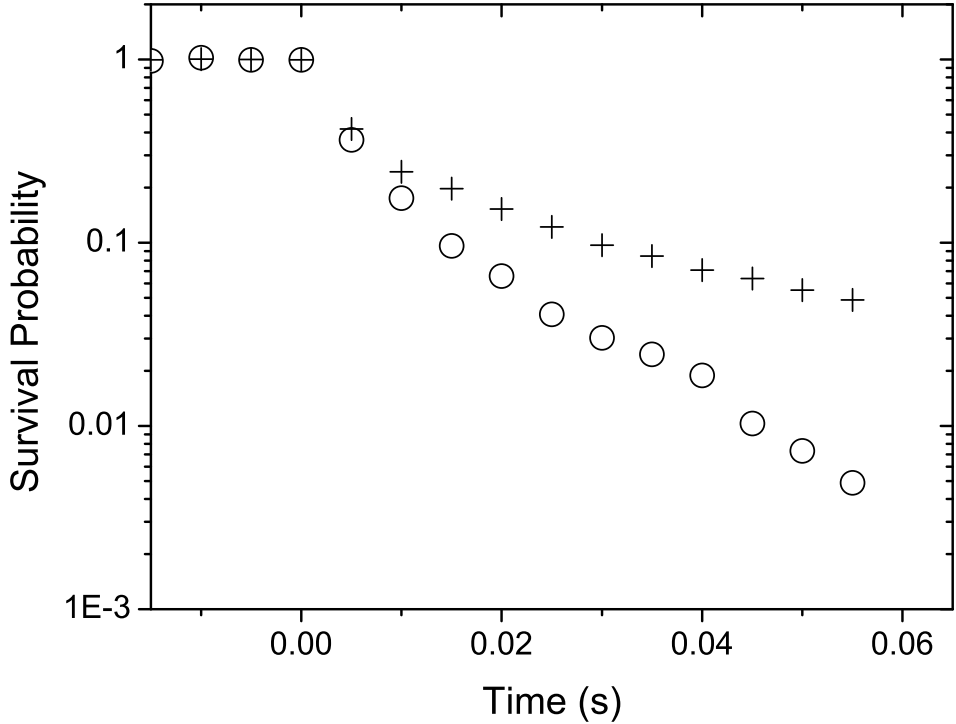}
\end{center}
\caption{Experimental results for the decay of cold atoms from a
tilted-stadium shaped atom-optics billiard, with two different
values for
the softness parameter: $w_{0}=14.5\protect\mu $m ($\circ $), and $w_{0}=24%
\protect\mu $m ($+$). The hole is located inside the big
semicircle. The smoothening of the potential wall causes a growth
in stability, and a slowing down in the decay curve. }
\label{expstad}
\end{figure}

\begin{figure}[tbp]
\begin{center}
\includegraphics[width=5in]{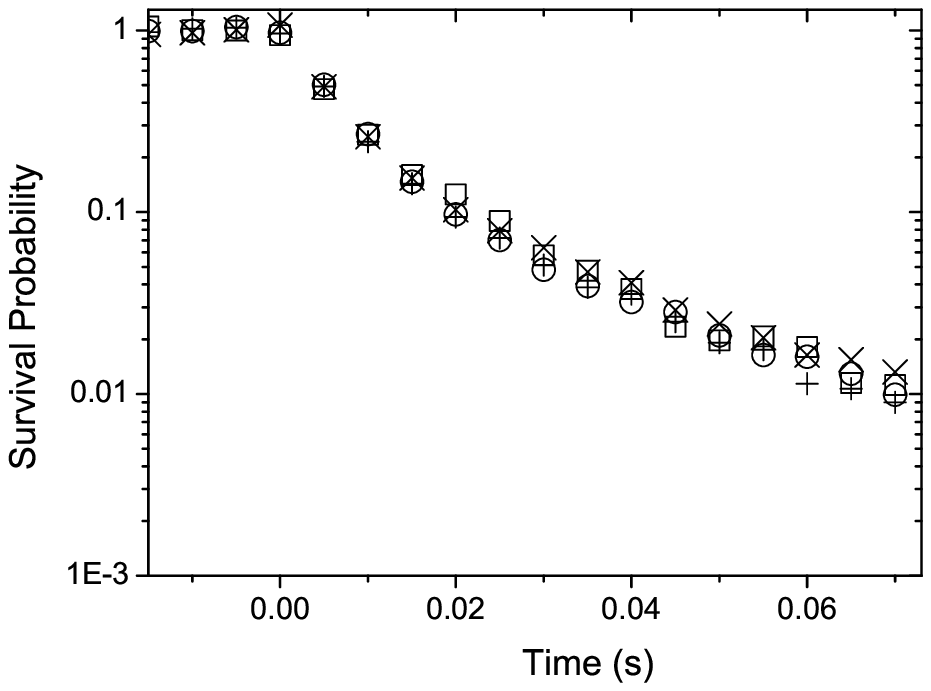}
\end{center}
\caption{Experimental results for the decay of cold atoms from a
circle shaped atom-optics billiard, with different values for the
softness parameter: $w_{0}=14.5\protect\mu $m ($\circ$),
$w_{0}=18.5\protect\mu $m (+), $w_{0}=21\protect\mu $m
($\square$), $w_{0}=23.5\protect\mu $m ($\times$). }
\label{expcirc}
\end{figure}

As a way to probe the nature of the dynamics of the atoms confined
by such an "atom-optics" billiard we measure the decay of the
number of confined atoms through a small hole on the boundary. A
purely exponential decay is expected for chaotic
motion\cite{Bauer90,Alt96,Kokshenev00}, with a decay time given by
$\tau _{c}=\pi A/vL$, where $A$ is the billiard's area, $v$ the
atomic velocity and $L$ the length of the hole \cite{Bauer90}. A
slower algebraic decay usually characterizes a stable or partly
stable phase space. In our system, the hole is produced by
switching off one of the AOSs for $\sim 1$ $\mu $s every scan
cycle, synchronously with the scan. The number of atoms remaining
in the trap is measured using fluorescence
detection\cite{friedman00}. The ratio of the number of trapped
atoms with and without the hole, as a function of time, is the
main data of our experiments.

First, we study a tilted Bunimovich stadium (see Fig.
\ref{trajstad}), which is chaotic and ergodic for the ideal
hard-wall case. We used a "tilted" stadium, to avoid the long
segments of regular motion associated with "bouncing ball"
trajectories in the original stadium\cite{Vivaldi83}. It is
composed of two semicircles of different radii ($64$ $\mu $m and
$31$ $\mu $m), connected by two non-parallel straight lines ($192$
$\mu $m long). The results of numerical simulations for the decay
in the number of confined atoms, through a small hole in the
boundary, are presented in Fig. \ref{simstad}. For the stadium
with $w_{0}=12$ $\protect\mu$m a nearly exponential decay is seen,
as expected for the ideal hard-wall billiard. As the walls are
made softer, a monotonic slowing down in the decay is observed. We
compare this behavior with that of a circle with the same area and
hole size, which exhibits integrable motion. In the ideal
hard-wall case the existence of nearly periodic trajectories
results in many time scales for the decay through a small hole on
the boundary, and then yield an algebraic decay\cite{Bauer90}.
Simulations for the decay from a circular billiard (dotted lines
in Fig. \ref{simstad}) show that the decay is almost unaffected by
the change in softness parameter, as expected.

Figure \ref{psstad} shows the Poincar\'{e} surface of section
obtained from the results of numerical simulations for classical
trajectories of Rb atoms inside the tilted-stadium billiard, with
$w_{0}=24\protect\mu $m. For clarity, we assume a monoenergetic
ensemble (with $v=120$ mm/s), a two-dimensional system, and no
gravity. However, including the experimental velocity spread,
three dimensional motion and gravity have only a small effect on
the results. In analogy to our experimental setup, we use for our
simulation a scanning gaussian potential, with a scan frequency of
100 kHz\cite{scanfreq}. The dimensions of the billiard are equal
to the experimental ones. The Poincar\'{e} surface of section
shows $ v_{x}$ versus $x$ at every trajectory intersection with
the billiard's symmetry axis ($y=0$), provided that $v_{y}>0$. The
smoothness of the wall results in a stability region that appears
around the singular trajectory which connects the points where the
big semicircle joins the straight lines. This fact is also
confirmed by analyzing some typical trajectories (see Fig.
\ref{trajstad}). Additional islands (smaller than the central one)
appear around it, and correspond to trajectories with higher
periodicity, like the one shown in Fig. \ref{trajstad}C. Around
these islands there is a large area of ''stickiness'', where the
trajectories spend a long time. Such a ''sticky'' trajectory is
presented in Fig. \ref{trajstad}D. The exact structure of the
island and its vicinity depends on the softness parameter $w_{0}$
in a sensitive way\cite{Kaplan02}. In general, the size of the
island and the stickiness region increase with the softness
parameter. The slowing down in the decay is attributed mostly to
the sticky trajectories, which in general cover a phase space
volume greater than that of the island itself (see Fig. 2 of
\cite{Kaplan02}).

In Fig. \ref{expstad}, experimental results for the decay from a
tilted stadium shaped atom-optics billiard are presented, for two
different values of the softness parameter ($w_{0}=14.5$ $\mu $m
and $ w_{0}=24$ $\mu $m). The experimental results confirm that
the soft wall causes an increased stability, and a slowing down in
the decay curve, in good agreement with the numerical results. We
also observed\cite{Kaplan02} that when the hole includes the
singular point where the semicircle meets the straight line, no
effect for the change in $w_{0}$ is seen. In this case the stable
island that is formed around the singular trajectory and the
region of "stickiness" around it (see Fig. \ref{trajstad}) are
destroyed by the hole. By moving the hole we can thus
experimentally map the location of the stable island.

Similar decay measurements for a circular atom-optics billiard
(see Fig. \ref{expcirc}) showed no dependence on $w_{0}$ in the
range $14.5-24$ $ \mu $m. We were not able to see any dependence
on the hole position either.

\section{Islands of stability in dispersive billiards}
We studied the effect of a soft potential wall also in a
dispersing billiard. The billiard that we studied was formed from
the intersection of four arcs with different radii and it had a
tangency, namely, a specific periodic orbit is tangent to the
billiard's boundary at one point (see Fig. \ref{trajbut}). In this
billiard, the periodic orbit had a period of two. More generally,
we consider a family of such billiards, with a variable parameter
$\zeta $, which is the vertical distance between the lower arc and
the two-periodic orbit. For $\zeta =0$ the periodic orbit is
tangent to the lower arc. For $\zeta>0$ the lower arc is below the
periodic orbit, while for $\zeta <0$ it is above and actually
intercepts the orbit. A billiard with a two-periodic tangent orbit
was chosen, to increase the size of the island. Every hard-wall
billiard of this family is fully chaotic. This specific billiard
shape was motivated by a theoretical study which predicts the
formation of an elliptic island around such a tangency
\cite{Romkedar99}.

\begin{figure}[tbp]
\begin{center}
\includegraphics[width=5in]{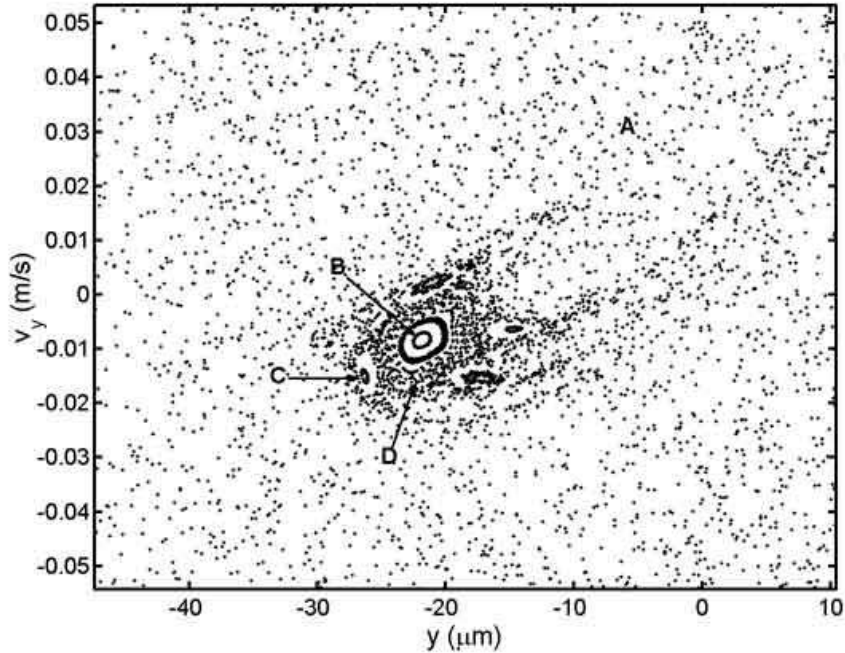}
\end{center}
\caption{Poincar\'{e} surface of section calculated for
monoenergetic atoms confined in the soft dispersing billiard
described in the text with $ \protect\zeta =40$ $\protect\mu $m
and $ w_{0}=24$ $\protect\mu $m. The Figure shows $\sim 17\%$ of
the total phase space area. Phase space area corresponding to
chaotic trajectories are marked "A", those of period 2
trajectories inside an island "B", areas of higher periodicity
trajectories "C" and "sticky" trajectories "D". } \label{psbut}
\end{figure}

\begin{figure}[tbp]
\begin{center}
\includegraphics[width=5in]{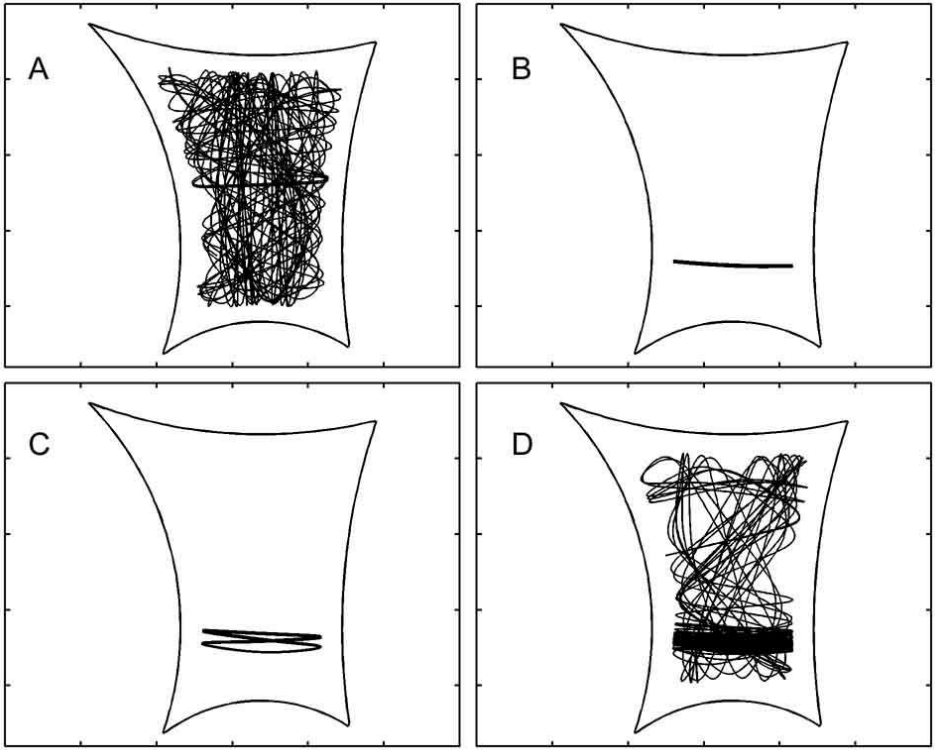}
\end{center}
\caption{Typical trajectories  in a soft dispersive billiard, with
the parameters as in Fig. \ref{psbut}, and softness parameter
$w_{0}=24$ $\mu $m. (A) A chaotic trajectory. (B) A periodic
trajectory of period 2. (C) A high periodicity trajectory. (D) A
"Sticky" trajectory.}\label{trajbut}
\end{figure}

\begin{figure}[tbp]
\begin{center}
\includegraphics[width=5in]{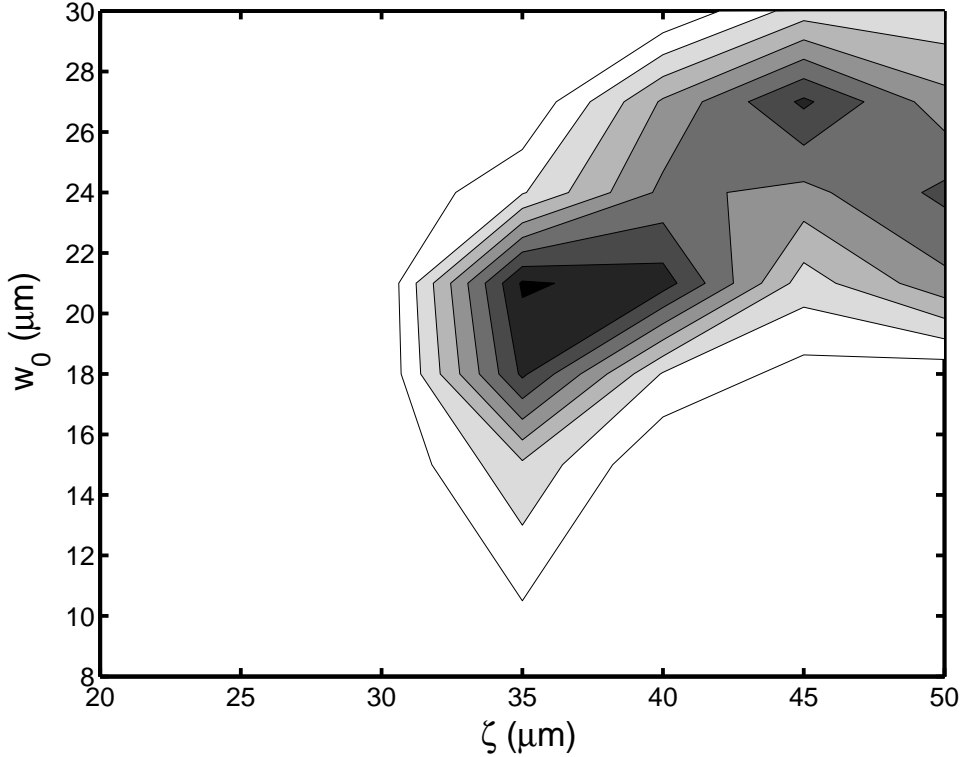}
\end{center}
\caption{Island size in the soft wall dispersing billiard, as a
function of the softness parameter, $w_{0}$, and the shape
parameter, $\protect\zeta $. (White - no island with an area
greater than $1.6\times10^{-3}$ of the total phase space area,
black - a large island, $\geq 4\times10^{-2}$ of the total phase
space).} \label{gogbut}
\end{figure}

In Fig. \ref{psbut}, the Poincar\'{e} surface of section in the
vicinity of the two-periodic orbit is presented, for a billiard
with $ \zeta =40$ $\mu $m, and $w_{0}=24$ $\mu $m. An elliptic
island is clearly observed. Typical trajectories for this case are
shown in Fig. \ref{trajbut}. In Fig. \ref{trajbut}A, a typical
chaotic trajectory is shown, corresponding to the areas in phase
space marked as "A" in Fig. \ref{psbut}.  A periodic trajectory of
period 2 (corresponding to the central island in Fig \ref{psbut},
marked as "B") is shown in Fig. \ref{trajbut}B. Also shown in Fig.
\ref{trajbut} are a high-period periodic trajectory and a "sticky"
one. We repeated these calculations for billiards with different
values of $\zeta $ and $w_{0}$, and plotted the size of the island
as a function of these two parameters in Fig. \ref{gogbut}. It can
be seen that the island becomes observable above some softness of
the potential, and that its size initially increases with an
increased softness. However, further softening reduces the size of
the island. In general, higher values of $\zeta $ require higher
values of $w_{0}$ to induce the stability. These results, although
obtained for a strongly perturbed case, are in agreement with
Refs. \cite{Turaev98,Romkedar99}. Some intuition can be gained by
noting that the stable trajectory is slightly reflected by the
lower arc. Since the stable trajectory is almost tangent to the
lower arc, its perpendicular velocity is very low, and a small
potential is already sufficient to reflect it. This small
potential is provided at a certain distance from the lower arc,
depending on $w_{0}$. If $\zeta $ is increased, the trajectory
will "feel" the lower arc (and hence will be stabilized) only for
a larger value of $w_{0}$. However, if $w_{0}$ is further
increased, the potential at the trajectory would become too high,
thus "blocking" it and destroying the stable island. In additional
simulations we observed that the center of the island moves
towards larger $ y$ values when $w_{0}$ is increased, in good
agreement with the above intuitive model.

\begin{figure}[tbp]
\begin{center}
\includegraphics[width=5in]{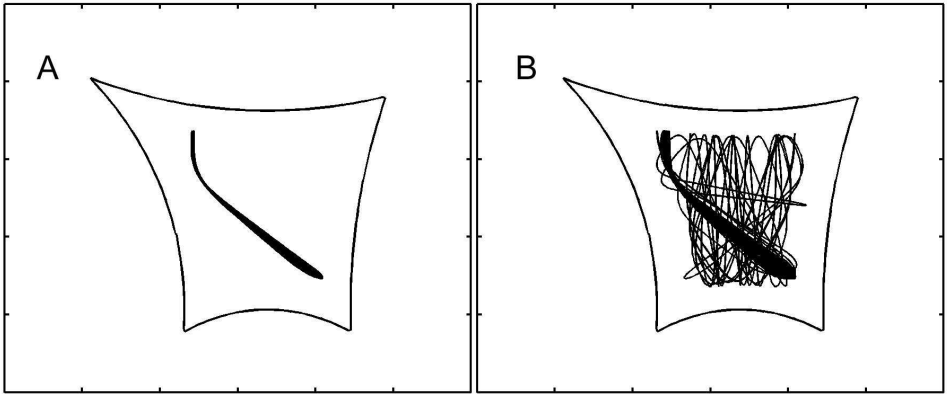}
\end{center}
\caption{Corner trajectories. (A) A periodic trajectory of period
2. (B) A "Sticky" trajectory.} \label{trajbut2}
\end{figure}

We observed also another kind of stable islands in these
billiards, around a different type of singular trajectories, i.e.
trajectories that are supported by the corners of the billiard. An
example for such a trajectory, and a sticky trajectory in the area
surrounding this new island, are shown in Fig. \ref{trajbut2}. Our
observations are consistent with the theoretical proof for the
existence of this large islands, provided in Ref. \cite{Turaev02}.
There it was shown that, for a small perturbation, its existence
depends both on the geometry of the corner and on the smooth
potential behavior at the corners. Our results show the existence
of this island also for a strong perturbation. In all cases where
an island was found, we also found a sticky region near it with
atoms spending arbitrary long times near the island.

We performed decay simulations for the scattering billiard, by
opening a hole on the boundary, and counting the number of
remaining atoms as a function of time. Here, as in the case of the
Stadium billiard, an exponential decay over 3 orders of magnitude
was observed for $w_{0}=8$ $\mu $m. In cases where an island
develops, a slowing in the decay was observed when the hole did
not include a reflection point of the periodic orbit. This slowing
down in the decay was similar to that for the stadium billiard of
Fig. \ref{simstad}.

\section{Summary}
In this paper we showed, numerically and experimentally, that
stable islands are created in the phase space of chaotic
billiards, either dispersive or focusing, when the walls are
softened. Using cold atoms in atom-optics billiards, an island of
stability was demonstrated in a tilted Bunimovich stadium
billiard, around the periodic trajectory that connects the points
where the big semicircle joins the straight lines. For a
dispersive billiard, we confirmed numerically the existence of a
stable island around a periodic orbit which is tangent to the
billiard's wall. In addition, we found numerical evidence, for the
first time, for the formation of a stable island near a "corner"
trajectory. In all three cases, the elliptic islands appear around
trajectories which are singular in the ''ideal'' infinitely hard
wall billiard. The orbits are singular in the sense that they do
not have a smooth limit when approaching the singularity from both
sides.

Our numerical results show that the appearance of a stable island
in a soft-wall billiard is in general accompanied by areas of
"stickiness" surrounding it. Although the size of the stable
island is a sensitive function of softness, not always showing a
monotonic behavior, we confirmed numerically and experimentally
that the "sticky" trajectories modify this behavior and, in
general, introduce a monotonic slowing down in the decay of the
number of trapped particles, as the walls are made softer.

An interesting question in this context is wether the existence of
a singularity in the hard-wall billiard is a necessary and
sufficient condition for stability in the soft-wall one. More
generally, one can ask to what extent is the relation between the
existence of a singularity in a chaotic system and the emergence
of stability when this system is perturbed, a universal feature.
From the practical point of view, the softness of the wall can be
used as a control parameter for the structure of phase space, in
particular in the context of controlling chaos\cite{Ott90}.
Finally, we note that since stable and "sticky" regions follow the
softness of the walls, and since any physically realizable
potential is inherently soft, the physical realization of a truly
chaotic and ergodic billiard is questioned.

\section{Acknoledgements}
We acknowledge helpful discussions with V. Rom-Kedar and support
from the Israel Science Foundation, the Minerva Foundation, the
United States-Israel Binational Science Foundation and Foundation
Antorchas. M. F. Andersen acknowledges support from the Nachemsohn
Dansk-Israelsk Studienfond.


\begin{thebibliography}{00}
\bibitem{scanfreq} We found our results to be independent of the
scanning rate, for rates $\geq$ 20 kHz.
\end{thebibliography}
\end{document}